\documentclass[12pt]{article}
\usepackage{cite}
\usepackage{color}
\usepackage{graphicx}
\usepackage{amsmath}
\usepackage{amssymb}

\makeatletter
\@addtoreset{equation}{section}

\makeatletter
\renewcommand\section{\@startsection {section}{1}{\z@}%
                                   {-3.5ex \@plus -1ex \@minus -.2ex}%nn
                                   {2.3ex \@plus.2ex}%
                                   {\normalfont\large\bfseries}}
\renewcommand\subsection{\@startsection{subsection}{2}{\z@}%
                                     {-3.25ex\@plus -1ex \@minus -.2ex}%
                                     {1.5ex \@plus .2ex}%
                                     {\normalfont\bfseries}}

\def\baselinestretch{1.2}
\newcommand{\be}{\begin{equation}}
\newcommand{\ee}{\end{equation}}
\newcommand{\beq}{\begin{eqnarray}}
\newcommand{\eeq}{\end{eqnarray}}

\newcommand{\tr}{{\rm Tr}}
\newcommand{\gone}[1]{{}}

\begin{document}
\begin{titlepage}
\begin{flushright}
hep-th/0512217\\
MAD-TH-05-08
\end{flushright}
%\vspace{12 mm}

\vfil\
%vfil

\begin{center}

{\Large{\bf Open/closed duality for FZZT branes in $c = 1$}}

\vfil

Ian Ellwood and Akikazu Hashimoto

\vfil

Department of Physics\\
University of Wisconsin\\ Madison, WI 53706\\

\vfil

\end{center}

%%%%%%%%%%%%%%%%%%%%%%%%%%%%%%%%%%%%%%%%%%%%%%%%%%%%%%%%%%%%%%%%%%%%%%%%%%%%%%%%%%%%%%%
\begin{abstract}
\noindent 
We describe how the matrix integral of Imbimbo and Mukhi arises from a
limit of the FZZT partition function in the double-scaled $c=1$ matrix
model.  We show a similar result for 0A and comment on subtleties in
0B. \end{abstract}
%%%%%%%%%%%%%%%%%%%%%%%%%%%%%%%%%%%%%%%%%%%%%%%%%%%%%%%%%%%%%%%%%%%%%%%%%%%%%%%%%%%%%%%%%
\vspace{0.5in}

\end{titlepage}
\renewcommand{\baselinestretch}{1.05}  %Line spacing
%%%%%%%%%%%%%%%%%%%%%%%%%%%%%%%%%%%%%%%%%%%%%%%%%%%%%%%%%%%%%%%%%%%%%%%%%%%%%%%%%%%%%%%%%%%%%

\section{Introduction}

Non-critical string theories with $c \le 1$ are an important
conceptual laboratory for exploring the inner workings of string
theory. As world sheet theories, they are formulated by coupling $c
\le 1$ matter to Liouville theory and describe strings propagating in
a space-time with a linear dilaton and a tachyon condensate.  While
these world sheet theories are, in general, as difficult to solve as
critical string theories, it is a special feature of low-dimensional
strings that there often exists a dual description in terms of an
exactly solvable matrix model
\cite{Klebanov:1991qa,Dijkgraaf:1991qh,Morozov:1994hh,Ginsparg:1993is,DiFrancesco:1993nw}.

There are, typically, {\em two} distinct formulations of a given $c\le
1$ string as a matrix model: the double-scaled matrix model and the
Kontsevich model. Historically, these models were derived in
completely different ways. The double-scaled matrix model is based on
`t Hooft's idea that, at large $N$, the Feynman diagrams of gauge
theory can be thought of as random triangulations of a string world
sheet.  The Kontsevich model, meanwhile, is found by directly solving
the flow equations implied by the integrable structure.  Amazingly,
this solution takes the form of a simple matrix integral
\cite{Kontsevich:1992ti}.  Both approaches can be used to compute
observables, such as correlation functions of vertex operators on the
world sheet, and give identical results. However, until recently, it
was not understood how they are related to each other.

The existence of dual formulations, one in terms of a string theory
and the other in terms of matrices, is reminiscent of
open/closed-string duality and suggests that the matrix models may
arise from open-string theories living on branes.  There are two
classes of branes in low dimensional string theories, which arise from
two possible boundary conditions in the Liouville sector: FZZT branes
\cite{Fateev:2000ik,Teschner:2000md}, which are the analogue of
Neumann boundary conditions and are extended, and ZZ branes
\cite{Zamolodchikov:2001ah}, which are the analogue of Dirichlet
boundary conditions and are localized in the strong coupling region.

ZZ branes are unstable.  This, together with the fact that they live
in the strongly coupled region, makes a direct construction of the
open string theory living on them difficult.  However, it was shown in
\cite{McGreevy:2003kb,Klebanov:2003km} that the decay of a ZZ brane
can be interpreted as a single eigenvalue of the double-scaled matrix
model rolling down the unstable potential. This allows one to think of
the double-scaled matrix model as the open string theory living on an
infinite number of ZZ-branes that have condensed to form a
Fermi-surface.

\begin{figure}
\centerline{
\begin{picture}(0,0)%
\includegraphics{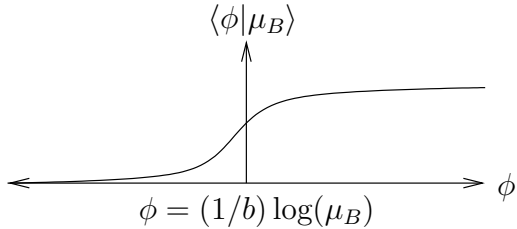}%
\end{picture}%
\setlength{\unitlength}{3947sp}%
\begingroup\makeatletter\ifx\SetFigFont\undefined%
\gdef\SetFigFont#1#2#3#4#5{%
  \reset@font\fontsize{#1}{#2pt}%
  \fontfamily{#3}\fontseries{#4}\fontshape{#5}%
  \selectfont}%
\fi\endgroup%
\begin{picture}(3602,1502)(900,-5162)
\put(2626,-3886){\makebox(0,0)[lb]{\smash{\SetFigFont{12}{14.4}{\rmdefault}{\mddefault}{\updefault}{\color[rgb]{0,0,0}$\!\!\!\!\! \langle \phi | \mu_B \rangle$}%
}}}
\put(4276,-4936){\makebox(0,0)[lb]{\smash{\SetFigFont{12}{14.4}{\rmdefault}{\mddefault}{\updefault}{\color[rgb]{0,0,0}$\phi$}%
}}}
\put(2026,-5086){\makebox(0,0)[lb]{\smash{\SetFigFont{12}{14.4}{\rmdefault}{\mddefault}{\updefault}{\color[rgb]{0,0,0}$\phi = (1/b) \log(\mu_B)$}%
}}}
\end{picture}}
\caption{Illustration of profile $\langle \phi | \mu_B \rangle \sim e^{- \mu_B e^{-b \phi}}$ of the boundary state $|\mu_B \rangle$ \label{figa}}
\end{figure}

FZZT branes are stable. They are labeled by a parameter $\mu_B$, known
as the boundary cosmological constant.  Although the FZZT brane is the
analogue of Neumann boundary conditions, it is only semi-infinite in
extent.  For example, if we denote by $|\mu_B \rangle$ the boundary
state of an FZZT-brane in the $(p,q)$ minimal model, then, in the
mini-superspace approximation, we have the profile\footnote{Our
convention is that $\phi \rightarrow +\infty$ is the weakly coupled
region.},
\be \langle \phi | \mu_B \rangle \sim e^{- \mu_B e^{-b \phi}}, \qquad
{\rm where} \qquad b^2 = {p \over q}, \qquad c = 1 - {6 (p-q)^2 \over
pq} \ . \ee
As one can see in figure \ref{figa}, the brane exists throughout the
weak coupling region at large $\phi$, but disappears around $\phi \sim
\frac{1}{b} \log \mu_B$.  Hence, by varying $\mu_B$, FZZT branes can
be used as a probe of target-space \cite{Seiberg:2003nm}.

Having identified the open string theory on ZZ branes with the
double-scaled matrix model, it is natural to try to identify the
Kontsevich model with open strings living on FZZT branes. The precise
relation was first pointed out by Gaiotto and Rastelli who considered
Witten's cubic open string field theory \cite{Witten:1985cc} for the
open strings ending on stack of $n$ FZZT branes for the case of the
minimal $(p,q) = (2,1)$ model \cite{Gaiotto:2003yb}.  Remarkably, they
found that the open string field theory \cite{Witten:1985cc} on a
stack of FZZT branes could be reduced to precisely the action of the
$n \times n$ Kontsevich matrix integral\footnote{Sometimes, the action
is presented in a form where the term linear in $M$ is replaced by
term quadratic in $M$ by shifting $M$.},
\be S(M,\Lambda) = \tr M \Lambda - \frac{M^3 }{ 3} \ . \label{konaction} \ee
%
%The fact that string field theory can reduce to an ordinary field
%theory is well known in the earlier example of the string field theory
%for the open strings in the topological A-model which reduces Chern
%Simons theory \cite{Witten:1992fb}. 
Unfortunately, while this result is elegant, it has not been easy to
generalize it to other examples, such as the $(p,q)\ne (2,1)$ models
\cite{Gaiotto:2003yb, Giusto:2004mt}.

An alternate approach, which seems to generalize more easily, is to
compute the open string theory on the FZZT branes using the double-scaled
matrix model.  An FZZT brane can be incorporated into the double-scaled
matrix model by inserting into the path integral the exponential of
the loop operator,
\be \exp(\tr \log(\mu_B - \Phi)) = \det(\mu_B - \Phi), \ee
where $\Phi$ is the $N \times N$ matrix of the double-scaled matrix
model.  This approach was pursued in \cite{Maldacena:2004sn}, and it
was shown that one could derive the Kontsevich model from the
partition function of the (2,1) matrix model in the presence of a
stack FZZT branes.

Unlike the string field theory analysis, the double-scaled matrix
model analysis can be generalized in a straightforward fashion to
other examples.  For example, it was shown in \cite{Hashimoto:2005bf}
that, using the the double-scaled $(p,1)$ model \cite{Daul:1993bg} in
the presence of $n$ FZZT-branes, one could derive the corresponding
generalization of the Kontsevich integral,
\be Z(\Lambda) = \int dM\,  
  e^{- \tr \left(M \Lambda - \frac{M^{p+1}}{ p+1}\right)} \  . \ee

In spite of the success of these arguments in $c<1$ models, the
extension to $c = 1$ has not been worked out.  As with the $c<1$
models there is an analogue of Kontsevich matrix integral,
\be Z_n(A,\bar t) = (\det A)^\nu \int dM\,  
   e^{\tr \left(-\nu MA 
    + (\nu-n) \log M 
    - \nu \sum_{k=1}^\infty \bar t_k M^k\right)}, 
\qquad 
   \nu = - i \mu\ ,  \label{imbimbo-mukhi}\ee
first derived by Imbimbo and Mukhi
\cite{Imbimbo:1995yv}.
However, the connection between the open string theory on 
FZZT-branes and the Imbimbo-Mukhi model has not been made.  

Several attempts to make this connection can be found in the
literature. One approach is to work out the string field theory living
on a stack of FZZT-branes in the same spirit as the Gaiotto-Rastelli
computation.  It is not clear, however, how the degrees of freedom of
open strings ending on a semi-infinite world volume should arrange
themselves into parameters $A$ and $\bar t$ of the Imbimbo-Mukhi model
and the authors of \cite{Ghoshal:2004mw} did not succeed in recovering
the Imbimbo-Mukhi model this way.

Another method of deriving the Imbimbo-Mukhi model used the so-called
normal-matrix model \cite{Alexandrov:2003qk}.  This model is similar
to the Imbimbo-Mukhi model in that it is a matrix integral solution to
the integrable flow equations.  As discussed in
\cite{Mukherjee:2005aq}, one can perform a computation similar to
\cite{Maldacena:2004sn,Hashimoto:2005bf} in which one evaluates the
expectation value of certain determinant operators in the model to
reproduce the Imbimbo-Mukhi model.  While this gives a nice connection
between the normal-matrix model and the Imbimbo-Mukhi model, the
relationship between the determinant operators in the normal-matrix
model and FZZT branes is not clear.

\section{FZZT branes in bosonic $c=1$ string theory}

Having reviewed what is known about the $c = 1$ case, we can now turn
to the main goal of this paper: we wish to fill in the conceptual gap
between the open string theory on the FZZT branes and the
Imbimbo-Mukhi model.  To do so, we attempt to compute the expectation
value of the exponential of the loop operator, as was done in the
$c<1$ case, and manipulate it into the form of the Imbimbo-Mukhi
matrix integral \cite{Hashimoto:2005bf}.  As we will see, many of the
ingredients that were used in the $c<1$ computation have analogues in
the $c = 1$ case.  There is, however, one new subtlety. As described
below, one must take a certain scaling limit of the FZZT partition
function before one can derive the Imbimbo-Mukhi model.

We begin our discussion with a brief review of the relevant aspects of
the $c = 1$ matrix model that we need for our derivation.  We mostly
follow the approach of Kostov \cite{Kostov:2002tk,KostovTalk}.  The
matrix-model action is given by
\begin{equation} \label{matrixmodelaction}
  S = \text{Tr} \int_0^\beta dx \,\Bigl(-\tfrac{1}{2} (\nabla_A \Phi)^2 +
  \tfrac{1}{2} \Phi^2 \Bigr),
\end{equation}
where 
\begin{equation}
   \nabla_A \Phi = \partial_x \Phi - i [A,\Phi],
\end{equation}
and $\Phi(x)$ and $A(x)$ are $N \times N$ Hermitian matrices.  For the
bosonic theory, we work in units where $\alpha'=1$. Since we are
interested in deriving the Imbimbo-Mukhi model, we work with the
Euclidean action and compactify the time direction with period $\beta
= 2\pi R$.

It is useful to rewrite the action, (\ref{matrixmodelaction}), in
chiral form.  To do this, one first switches to the first order form,
\be S = \tr
\int_0^\beta dx \left(i P(\nabla_A \Phi) - \tfrac{1}{2} P^2 +
\tfrac{1}{2} \Phi^2 \right), \ee
which reduces to equation (\ref{matrixmodelaction}) upon integrating
out $P$. Then, substituting
\begin{equation}
  X_{\pm} = \frac{\Phi \pm P}{\sqrt{2}},
\end{equation}
one arrives at the chiral action,
\be S = \tr \int_0^\beta dx ( - i X_+ \nabla_A X_- + X_+ X_- )
\label{chiral}. \ee
In this form it is relatively straightforward to reduce the path
integral over $X_\pm$ and $A$ to an ordinary three-matrix integral
\cite{Alexandrov:2002fh,KostovTalk,Kostov:2002tk}.  Defining $X_\pm = X_{\pm} (t
= 0)$ one finds,
\be Z_N = \int dX_+ dX_- d \Omega e^{i \tr(X_+ X_- - q^{-1} X_+ \Omega
X_- \Omega^{-1})} \label{3matrix},\ee
where the subscript, $N$, is a reminder that the partition function
depends on the rank, $N$, of the matrices and $q = e^{i\beta}$. The
matrix, $\Omega \in U(N)$, is given by
\be \Omega = P \left[ e^{i \int_0^\beta A(x) dx} \right]. \ee
This expression can be simplified
further by integrating out $\Omega$, which reduces the partition function
to an integral over the eigenvalues of $X_{\pm}$ ,
\be Z_N = \int \prod_{k=1}^N d x_k^+ dx_k^- \det_{jk}(e^{-i x_j^-
x_k^+}) \det_{jk} (e^{i q^{-1} x_j^- x_k^+}) \label{singlet}. \ee
In order to extract the observables of the $c=1$ matrix model, one
computes the grand partition function, which, following
\cite{Alexandrov:2002fh,KostovTalk}, can be expressed in the form of a
Fredholm determinant,
\be Z(\mu) = e^{-\beta \mu N} Z_N = \det(1 + e^{-\beta \mu} K_- K_+), \ee
where
\begin{align}
 [K_+ f](x_-) &= \int dx_+ e^{i q^{-1} x_+ x_-} f(x_+),  
\nonumber\\
[K_- f](x_+) &= \int dx_- e^{-i  x_+ x_-} f(x_-)  
\label{measurefactor}
\end{align}
are operators acting on functions of $x_+$ and $x_-$,
respectively. The Fredholm determinant is computed by diagonalizing
the operators $K_+$ and $K_-$ and taking the product of their
eigenvalues. This closely resembles what one does in computing the
partition for the two-matrix models in terms of orthogonal
polynomials.  Up to non-perturbative corrections, the partition
function of the undeformed Euclidean $c=1$ model can be written as
\be \log Z(\mu) = \sum_{r = Z + {1 \over 2}} i \phi_0 (-\mu + {i \over
R} r), \qquad e^{i \phi_0(E)} = e^{i \pi /4} {\cosh(\pi E) \over \pi}
\Gamma\left({1 \over 2} + i E\right) \label{initial}. \ee

As is well known, when compactified on a circle, the $c = 1$ theory
contains a tower of tachyon deformations with discrete Euclidean
momenta.  These deformations change the asymptotic shape of the Fermi
surface in phase space. One can incorporate them into the matrix model
by generalizing (\ref{singlet}) to
\be Z_N = \int \prod_{k=1}^N [d x_k^+] [dx_k^-] \det_{jk}(e^{-i x_j^-
x_k^+}) \det_{jk} (e^{i q^{-1} x_j^- x_k^+}) \label{singlet2}, \ee
where
\be 
[dx_\pm] = e^{\pm i U_\pm(x_\pm)} dx_\pm, \qquad U_\pm(x_\pm) = i
z_{\pm k} x_\pm^{k/R} \label{deform}\ee
and the $z_k$ are the deformation parameters corresponding to tachyons
with momentum $k$. The sign and the phase of $z_k$ was chosen so that
it coincides with the deformation parameter of
\cite{TakasakiUeno}. This identification is explained in more detail
in appendix A.  The grand partition function is still given by the
Fredholm determinant, but for the deformed operator,
\be Z(\mu) = \det (1 + e^{\beta \mu} \tilde{K}_- \tilde K_+),\ee
where
\begin{align}
\label{deformedK}
[\tilde K_+ f](x_-) &= \int [dx_+] e^{i
q^{-1} x_+ x_-} f(x_+),
\nonumber\\
[\tilde K_- f](x_+) &= \int [dx_-] e^{-i
 x_+ x_-} f(x_-) \ . 
\end{align}
It was shown in \cite{Kostov:2002tk} that such a deformation can be
described in terms of a Toda flow and that the partition function is
related to the $\tau$-function via\footnote{The connection between
$c=1$ matrix model and the Toda integrable hierarchy was first pointed
out in \cite{Dijkgraaf:1992hk}. For introduction to Toda lattice
hierarchy, see \cite{TakasakiUeno,Takebe:1990fz,TakasakiBook}}
\be Z(\mu) = \tau\left( R(i \mu-\tfrac{1}{2} ) +\tfrac{1}{2}, z_+,
z_-\right) \ .  \ee
The initial condition for the flow, (\ref{initial}), can also be
expressed in the form of the string equation, found in
\cite{Takasaki:1995fu,Kostov:2001wv}.

Having given a brief review of the $c = 1$ matrix model, we now study what
happens when we add an FZZT-brane to the story.  As
mentioned earlier, an FZZT-brane can be incorporated in the matrix
model by inserting into the path integral an exponential of the
macroscopic-loop operator,
\be \label{fzztdet}
e^{\pm W(x)} = \det (\Phi(x) - \mu_B)^{\pm 1} = \det \Bigl((X_+(x) +
X_-(x))/\sqrt{2} - \mu_B\Bigr)^{\pm 1}, \ee
as first considered in \cite{Moore:1991sf}.  Note that the determinant
operator generates an FZZT brane with Dirichlet boundary conditions in
the $c = 1$ direction and, hence, is localized in Euclidean time at a
point $x$.  The ``$\pm$'' in the exponent is included so that we can
simultaneously consider the determinant and the inverse determinant
operator insertions.  As we will see later, the connection with the
Imbimbo-Mukhi model favors the inverse determinant
\cite{Mukherjee:2005aq}.  However, inserting the determinant also
gives rise to a matrix integral that encodes the $\tau$-function of
the Toda integrable flow and is, roughly, the complex conjugate of the
Imbimbo-Mukhi model.

Inserting (\ref{fzztdet}) into the chiral Lagrangian,
(\ref{chiral}), one sees that the determinant has a simple $x$
dependence,
\be e^{\pm W(x)} = \det\Bigl((e^{i x/R} X_+(t = 0) + e^{-i
x/R} X_-(t = 0)) / \sqrt{2} - \mu_B\Bigr)^{\pm}. \ee
One can also see this $x$ dependence by considering the Heisenberg form 
of the operators, $X_\pm$, in the Hamiltonian language.

Now, it would be nice if inserting this operator into the path
integral was equivalent to a tachyon deformation.  It clearly
is not, since the determinant contains a complicated mixture of $X_+$
and $X_-$ operators, and  does not factorize into an $X_+$ term
and an $X_-$ term in the way that the tachyon deformation,
(\ref{deform}), does.  However, as we now show, it  is possible
to get a pure tachyon deformation if we take a certain limit of the
determinant operator.

Suppose that we analytically continue $x = -it$ and send $t \to
\infty$ while simultaneously rescaling
\be \mu_B = e^{t} \mu_B' \ . \label{scaling}\ee
Then, up to an overall normalization factor, the operator asymptotes to
\be \det\left(1 - {X_+ \over \mu_B'} \right)^{\pm 1} = e^{
\mp \sum_n {1 \over n} \mu_B'^{-n} X_+^n} \
. \label{scaledloop} \ee
In the case where the radius of the $c=1$ direction is one, inserting
this operator into the partition function is equivalent to a shift in
the Toda flow,
\be \label{TodaFlow} \delta z_k = \pm  {1 \over n} \mu_B'^{-n},
\qquad \delta k >0.
\ee
Readers familiar with \cite{Mukherjee:2005aq} will recognize this
deformation as the same one generated by one of the determinants in
the normal-matrix model, $\det ( 1 - Z /\mu_B)^{\pm 1}$.  The other
determinant, $\det ( 1 - Z^\dagger /\mu_B)^{\pm 1}$, can generated by
taking the opposite sign in the analytic continuation, $x \to i t$ .
Notice that away from $R = 1$ there does not seem to be any relation
between the FZZT-brane determinant and the determinant of the normal
matrix model.

Once we have identified that our limit of the FZZT determinant
generates the Toda flow, (\ref{TodaFlow}), the integrable structure of
the model is enough to determine that the expectation value of
(\ref{scaledloop}) will give the Imbimbo-Mukhi model, as was shown in
\cite{Mukherjee:2005aq} using the normal-matrix model.  However, we
would like to understand how to derive the Imbimbo-Mukhi model more
directly.

As in the $c<1$ case, it is useful to introduce the Baker-Akhiezer
function, $\Psi^{\pm}_s(\lambda)$, which is given in terms of the
$\tau$-function by
\begin{align}
\Psi^+_s(\lambda) &= {\tau( z_+ - \epsilon(\lambda^{-1/R}), z_-,s) \over 
\tau( z_+ , z_-,s)} \exp[z_+(\lambda^{1/R}) + s \log(\lambda^{1/R})] ,
\cr
\Psi^-_s(\lambda) & = {\tau(z_+,z_- - \epsilon(\lambda^{1/R}),s+1)
\over \tau(z_+, z_-,s)} \exp[ z_-(\lambda^{-1/R}) + s
\log(\lambda^{1/R}) ],
 \label{BA} 
\end{align}
where
\be \epsilon(\lambda) = \left(\lambda, {\lambda^2 \over 2}, {\lambda^3
\over 3}, \ldots \right), \qquad z_+(\lambda) = \sum_{n=1}^\infty z_n
\lambda^n, \qquad z_-(\lambda^{-1}) = \sum_{n=1}^\infty z_{-n}
\lambda^{-n} \ .  \ee
For a formal definition of the Baker-Akhiezer function in terms of
dressing operators in the Lax formalism, see equations (25) and (27)
of \cite{Takasaki:1993at}.

Notice that, in the expression for $\Psi^+_s$, the argument of the
$\tau$-function is shifted by $z_+ \to z_+ -
\epsilon(\lambda^{-1/R})$.  Setting $\lambda = \mu_B'$ and $R = 1$, we
see that this is the same shift as (\ref{TodaFlow}), provided we
consider the {\em inverse} determinant.  Hence, the $\tau$-function in the
presence of the inverse determinant, (\ref{scaledloop}), can be
computed if we know the Baker-Akhiezer function.

By itself, this fact is not very useful.  However, the Baker-Akhiezer
functions are also closely related to the biorthogonal eigenfunctions,
$\psi_\pm$, of the deformed operators, $\tilde{K}_\pm$ given in
(\ref{deformedK}) \cite{Kostov:2002tk}:
\be \psi_+^E(x_+)^* = \left. \Psi^+_s(x_+) \right|_{s = -R(iE + {1/
2})}, \qquad \psi_-^E(x_-)^* = \left. x_- \Psi^-_s(x_-^{-1})\right|_{s = -R
\left(iE + {1/ 2} \right) }.\ee
Since the biorthogonal eigenfunctions are related to each other by
Fourier transform, it follows that $\Psi^+_s$ and $\Psi^-_s$ are related
by
\be \Psi^+_s(\lambda_+) = \int d \lambda_-^{-1} e^{i \lambda_+
\lambda_-^{-1}} \lambda_- \Psi^{-}_s(\lambda_-).  \ee
Now, suppose that $z_+=0$.  Then the non-trivial pieces of $\Psi^-_s$
drop out and $\Psi_s^-$ is completely determined as a function of $z_-$
and $\lambda$;
\be \Psi^-_s(\lambda_-) = \mathcal{N} \exp[z_-(\lambda_-^{-1/R})+s
\log(\lambda_-^{1/R})], \ee
where $\mathcal{N}$ is a normalization factor independent of $\lambda$
and $z_-$, which we can ignore. Changing variables to $m =
\lambda_-^{-1}$, one has
\be \Psi^+_s(a) = \int dm\, \exp[ i a m + z_-(m^{1/R}) - (s+R) \log
(m^{1/R})]. \ee
Using (\ref{BA}), one finds
\be \tau(-\epsilon(a^{-1/R}),z_-,s) = \int dm\, (ma)^{-s/R} \exp[iam +
z_-(m^{1/R}) - \log(m)], \label{tau1} \ee
which is equivalent to
\be Z_1(A,t_-) = \tau(-\epsilon(a^{-1/R}),z_-,s)= \int dM\, (M A)^{\nu R
+ (R-1)/2} e^{- \nu ((M A)^R + t_-(M)) - \log(M)}, \ee
with the identifications,
\be 
\label{IMnotation}
t_{-n} = i \mu^{n/2R-1} z_{-n},\qquad m = \mu^{1/2} M^R, \qquad a
= \mu^{1/2} A^R, \qquad s = i \mu R - {R -1 \over 2}, \qquad \nu = -i
\mu. \ee 
As we discuss in appendix C, this argument can be
iterated to give the general expression,
\be Z_n(A,t_-) = \int dM_i \, {\Delta(M)
\over \Delta(A)} \prod_{i=1}^n \left[ (M_i A_i)^{\nu R + (R-1)/2} e^{-
\nu ((M_i A_i)^R + {1 \over n} t_{-n} M_i^n ) - n \log(M_i)}\right] \ ,
\label{notselfdual} \ee
with Miwa-Kontsevich-like transform,
\be t_n = - {1 \over \nu k} A^{-k} \ . \ee
This is the same result obtained in \cite{Mukherjee:2005aq} using the
normal-matrix model.\footnote{It should be emphasized that while the
sign of $z_n$'s is fixed by matching to the conventions of
\cite{TakasakiUeno}, there is no intrinsic definition for the sign of the
$t_n$'s. In particular, the sign convention for the $t_n$'s in
\cite{Imbimbo:1995yv} and \cite{Mukherjee:2005aq} appear to differ by
a sign.}  Setting $ R = 1$, we reproduce precisely the Imbimbo-Mukhi
integral. 

Recall that, to match with the argument of the Baker-Akhiezer
function, we had to consider the inverse determinant.  This is in
agreement with the observation made in \cite{Mukherjee:2005aq} that
the Imbimbo-Mukhi integral is derived from the insertion of an inverse
determinant in the normal-matrix model.

One can also consider the insertion of a determinant.  In this case,
one must work with the dual Baker-Akhiezer
functions \cite{TakasakiBook}\footnote{They are related to
$w^{\left({\infty \atop 0}\right)*}(s,z_+,z_-,\lambda)$ in the
notation of \cite{TakasakiUeno} according to
$$
 \Psi^{+*}_s(\lambda) = w^{(\infty)*}(s,z_+, z_-, \lambda)e^{-z_+(\lambda^{1/R})}\lambda^{-s/R} , \qquad
\Psi^{-*}_s(\lambda) = w^{(0)*}(s,z_+, z_-, \lambda) e^{-z_-(\lambda^{1/R})}\lambda^{-s/R} \ .$$},
\begin{align}
\Psi^{+*}_s(\lambda) &= {\tau( z_+ + \epsilon(\lambda^{-1/R}), z_-,s) \over 
\tau( z_+ , z_-,s)} \exp[-z_+(\lambda^{1/R}) - s \log(\lambda^{1/R})] ,
\cr
\Psi^{-*}_s(\lambda) & = {\tau(z_+,z_- + \epsilon(\lambda^{1/R}),s-1)
\over \tau(z_+, z_-,s)} \exp[ -z_-(\lambda^{-1/R}) - s
\log(\lambda^{1/R}) ]\ . 
 \label{BAdual} 
\end{align}
As with the Baker-Akhiezer functions, they are related by Fourier
transform;
\be  {1 \over \sqrt{2 \pi} }\Psi^{+*}_s(\lambda_+) = \lambda_+^{1-1/R} \int d\lambda_-^{-1}\,  e^{-i \lambda_+ \lambda_-^{-1}  }   {1 \over \sqrt{2 \pi} } \lambda_-^{1/R} \Psi^{-*}_s(\lambda_-), \label{fourier}
\ee
from which one can derive,
\beq Z_n(A,t_-) &=& \int dM_i \, {\Delta(M)
\over \Delta(A)} \prod_{i=1}^n \left[ (M_i A_i)^{\nu R + (R-1)/2} e^{
-\nu ((M_i A_i)^R + {1 \over n} t_{-n} M_i^n ) - n \log(M_i)}\right] 
\ , \cr
t_n & = &  -{1 \over \nu k} A^{-k}, \qquad \nu = i \mu \ . \label{notselfdual2} \eeq
For $R=1$, one can use the Harish-Chandra/Itzykson-Zuber formula
\cite{HarishChandra,Itzykson:1992ya} to re-express this integral in
matrix form. This again yields a matrix model identical to the
Imbimbo-Mukhi model.  The only difference between (\ref{notselfdual2}) and
(\ref{notselfdual}) is that we have changed the relative sign between
$\nu$ and $i \mu$.  Since both models are an even function of $\mu$,
this does not have any effect on the closed string
observables. Therefore, at the level of reproducing the
$\tau$-function, one can not distinguish between the determinant and
the inverse determinant. This choice is ultimately tied to whether the
FZZT-ZZ strings are bosons or fermions.

When $R \ne 1$ it is still true that (\ref{notselfdual}) and
(\ref{notselfdual2}) give complete information about the
$\tau$-function.  However, since the FZZT determinants no longer give
rise to a Toda flow, (\ref{notselfdual}) and (\ref{notselfdual2}) are
no longer related -- in any obvious way -- to FZZT-branes.  Thus, we
find that the open/closed string duality, relating the Kontsevich-like
matrix integral and the FZZT brane, appears to work only at self dual
radius. This is consistent with the observation in
\cite{Mukherjee:2005aq} that (\ref{notselfdual}) and
(\ref{notselfdual2}) cannot be represented in the form of a matrix
integral except at $R = 1$.  After all, if (\ref{notselfdual}) and
(\ref{notselfdual2}) were derivable from an open/closed duality, one
would expect the result to take the form of a matrix integral for
general $R$.

\section{FZZT branes in $\hat c=1$ string theories}

Our analysis relating the determinant operator to a Kontsevich-like
matrix integral can be extended to the $\hat c=1$ models
\cite{Takayanagi:2003sm,Douglas:2003up,Klebanov:2003wg}.  There are
two $\hat c=1$ models: type 0A and type 0B. We consider them
separately.

\subsection{Type 0A}

The type 0A theory is formulated in terms of a $U(N) \times U(N+q)$
gauged matrix model \cite{Douglas:2003up,Klebanov:2003wg}.  The
parameter $q$ gives the number of units of Ramond-Ramond flux in the
background.\footnote{See \cite{Maldacena:2005he} for a discussion
of the subtleties associated with the two ways of introducing
charges in this model.} As in the bosonic case, one can fix the gauge
and reduce the matrix degrees of freedom to the eigenvalues. This
gives rise to a model which is equivalent to the deformed matrix model
originally constructed in \cite{Jevicki:1993zg}.  We will work in
units where $\alpha' = 1/2$ so that the harmonic oscillator has the
same oscillator frequency as the $c=1$ model. Physical observables
such as the S-matrix and the partition function of this model were
computed in \cite{Demeterfi:1993cm}. These computations were then
reformulated in light-cone variables \cite{Maldacena:2005he}, which
are better suited for the analyzing the deformations and Toda
integrable structure of the model \cite{Yin:2003iv}.\footnote{See also
\cite{Johnson:2003hy,Seiberg:2004ei,Johnson:2004ut,Carlisle:2005wa}
for a related discussion in the context of minimal models.}

The main difference between the type 0A model and the bosonic $c=1$
model is that, in 0A, the wave functions, $\psi_+^E(x_+)$ and
$\psi_-^E(x_-)$, are related by a Bessel transform instead of a
Fourier transform. As a result, the Baker-Akhiezer functions are also
related by an integral Bessel transform,
\be \Psi^+_s(\lambda_+) = 
 \int_0^\infty d\lambda^{-1} _- \, 
 \sqrt{\lambda_+ \lambda_-^{-1}} J_q(\lambda_+ \lambda^{-1}_-) 
 \lambda_- \Psi^-_s(\lambda_-).
  \label{besseltransform}
\ee
The partition function can be written in the form\footnote{Note that
$q$ here refers to the background flux and not $e^{i \beta}$.},
\be Z_N = 
  \int \prod_{k=1}^N [d x_k^+] [dx_k^-] 
\det_{jk}\left(\sqrt{x_j^- x_k^+} J_q( x_j^- x_k^+) \right)
\det_{jk}\left(\sqrt{x_j^- x_k^+}J_q(e^{-i \beta} x_j^- x_k^+)\right) 
\label{singlet3}, \ee
where
\be 
[dx_\pm] = e^{\pm i U_\pm(x_\pm)} dx_\pm, \qquad U_\pm(x_\pm) = i
z_{\pm k} x_\pm^{k/R} \label{deform2}.
\ee
We can also express $Z(\mu)$ as a Fredholm determinant,
\be Z(\mu) = \det (1 + e^{\beta \mu} \tilde{K}_- \tilde K_+),\ee
where
\begin{align}
  [\tilde K_+ f](x_-) 
    &= 
   \int [dx_+]  
     \sqrt{x_+ x_-} J_q(e^{-i \beta} x_+ x_-) f(x_+)\ ,
\nonumber
\\
 [\tilde K_- f](x_+) 
    &= 
   \int [dx_-]
     \sqrt{x_+ x_-} J_q(- x_+ x_-) f(x_-)\ .
\label{deformedK2}
\end{align}

Using the relation between the $\tau$-function and the Baker-Akhiezer
function, (\ref{BA}), one finds
\be \tau( \mp \epsilon(a^{-1/R}), z_-,s) = \int d m (m a)^{-s/R+1/2}
J_q(am)\exp[ \pm z_-(m^{1/R}) - \log(m)], \ee
which generalizes (\ref{tau1}) to type 0A. Just as in the $c=1$ case,
one can rescale
\be  t_{-n}  = i \mu^{n/2R -1}z_{-n} , \qquad 
m = \mu^{1/2} M^R, \qquad a = \mu^{1/2} A^R
\ee
to bring this expression into an Imbimbo-Mukhi form.  Generalizing to the
multiple FZZT case one finds the expression,
\begin{multline}
\lefteqn{\tau( \mp  \epsilon(a^{-1/R}), z_-,s)} \\
 =
 \int d M \,   
  {\Delta(M) \over \Delta (A)} 
  \prod_{i=1}^n  (M_i A_i)^{-s+R/2}  
  J_q(\pm \mu A_i^R M_i^R)
  \exp[ \pm  i \mu     (  t_-(M_i)) -   n \log(M_i)],
 \label{0aim} 
\end{multline}
where 
\be 
s  =  \pm i \mu R - {R-1 \over 2}\ . \ee
Using
\be \int d \theta\,  e^{i q \theta - \nu  m a \cos \theta }  
   =  J_q(m) \ , \ee
one can integrate in an additional variable and express the shifted
$\tau$-function in a two-integral form:
\begin{multline}
\lefteqn{\tau( \mp \epsilon(a^{-1/R}), z_-,s)
 =  {1 \over \Delta(z)^{1/2R}}   
 \prod_{i=1}^n \left(z_i ^{ \nu/2  + (R-1/2)/4R +q/2}\right) 
 \int_0^\infty d s_i \, \prod_{i=1}^n  
  \left({1 \over s_i} e^{- s_i z_i+q}\right) }
\\
\times
 \int d y_i \, \Delta(y)^{1/2R} \,   
  \prod_{i=1}^n  y_i^{ \nu/2  + (R-1/2)/4R - (n+q/2)}  
  e^{ - y_i /s_i } \exp[  -\nu (  t_-(y_i^{1/2R})) ]\ ,    
\end{multline}
where we have used the change of variables,
\be y_i 
  = m_i^2\ , \qquad s = {m \over  a \nu} \cos \theta\  , 
 \qquad  
z_i = \nu^2 a_i^2\ , \qquad \nu = \mp i \mu \ . 
\ee
Setting $R=1/2$, one recovers a matrix integral representation found
in \cite{Ita:2004yn,Hyun:2005fq}.

To relate the Baker-Akhiezer function and the $\tau$-function to FZZT
branes, we consider the determinant operator \cite{Klebanov:2003wg},
\be 
e^{\pm W(x)} = \det (X(x) X(x)^\dag - \mu_B)^{\pm 1} = \det \left((X_+(x) +
X_-(x))(X_+^\dag(x) +
X_-^\dag(x))/2 - \mu_B\right)^{\pm 1}.
\label{0Afzzt}
\ee
Taking the same scaling limit as we did in the bosonic case,
(\ref{0Afzzt}) reduces to
\be \det\left(1 - {X_+ X_+^\dag  \over \mu_B'} \right)^{\pm 1} = e^{
\mp \sum_n {1 \over n} \mu_B'^{-n} (X_+ X_+^\dag)^n} \ . 
\label{scaledloop2} \ee
Comparing this with the parameterization of the deformation parameter,
(\ref{deform}), we see that the exponentiated macroscopic-loop operator
matches with the Baker-Akhiezer function at $R = 1/2$. This provides
the link between FZZT branes and the Imbimbo-Mukhi integral,
(\ref{0aim}), as well as the matrix integral of
\cite{Ita:2004yn,Hyun:2005fq}.

Since $\hat c=1$ models are on firmer non-perturbative footing that
the bosonic $c=1$ models, it is instructive to examine the the
deformed $\tau$-function function at the non-perturbative level.  To
that end, it is useful to go back to the integral transform,
(\ref{besseltransform}), and write it in the form,
\be 
\Psi^s_+(\lambda_+) = \int_0^\infty d\lambda_-^{-1} \,  
\sqrt{\lambda_+ \lambda^{-1}_-} \left( 
H^+_q(\lambda_+ \lambda^{-1}_-) + H^-_q(\lambda_+ \lambda^{-1}_-) \right)
\Psi_-^s (\lambda_-)  ,
\ee
where
\be H_q^\pm(x) = 
{J_q(x_+ x_-)  \pm i Y_q(x_+ x_-) \over 2} \ee
are the Henkel functions. 
Assuming that the background is only deformed by $z_-$, one can use
the relation between the Baker-Akhiezer function and the $\tau$-function
to write
\be \tau( - \epsilon(a^{-1/R}), z_-,s) = \int_0^\infty d m (m a)^{-s/R+1/2}
(H^+_q(am)+H^-_q(am))\exp[ z_-(m^{1/R}) - \log(m)] \label{integral1}\ . \ee
To define the theory non-perturbatively, one must specify the contour
of integration for the variable, $m$.  For the undeformed background,
$z_-=0$, one can integrate the Henkel function $H^\pm_q(x)$ anywhere
along the upper/lower half of the complex $x$ plane. This choice of
contour reproduces the correct perturbative expansions in
$z_-$.

\begin{figure}
\centerline{
\begin{picture}(0,0)%
\includegraphics{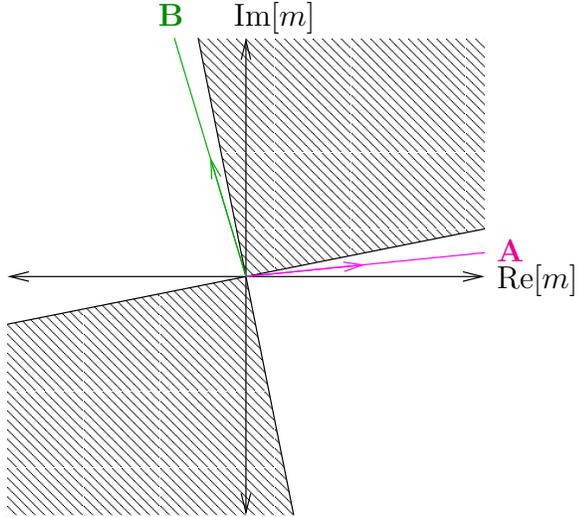}%
\end{picture}%
\setlength{\unitlength}{3947sp}%
\begingroup\makeatletter\ifx\SetFigFont\undefined%
\gdef\SetFigFont#1#2#3#4#5{%
  \reset@font\fontsize{#1}{#2pt}%
  \fontfamily{#3}\fontseries{#4}\fontshape{#5}%
  \selectfont}%
\fi\endgroup%
\begin{picture}(3602,3602)(2100,-3662)
\put(3350,-286){\makebox(0,0)[lb]{\smash{\SetFigFont{12}{14.4}{\rmdefault}{\mddefault}{\updefault}{\color[rgb]{0,0.55,0}{\bf B}}%
}}}
\put(5476,-1761){\makebox(0,0)[lb]{\smash{\SetFigFont{12}{14.4}{\rmdefault}{\mddefault}{\updefault}{\color{magenta}{\bf A}}%
}}}
\put(3826,-286){\makebox(0,0)[lb]{\smash{\SetFigFont{12}{14.4}{\rmdefault}{\mddefault}{\updefault}{\color[rgb]{0,0,0}Im[$m$]}%
}}}
\put(5476,-1936){\makebox(0,0)[lb]{\smash{\SetFigFont{12}{14.4}{\rmdefault}{\mddefault}{\updefault}{\color[rgb]{0,0,0}Re[$m$]}%
}}}
\end{picture}
}
\caption{An illustration of possible contours of integration in
$m$-space for which the integral (\ref{integral1}) converges. Contours
asymptoting to infinity along the shaded regions are excluded. Here it
is assumed that $z_{-1}$ is the only non-vanishing deformation
parameter.
\label{figb}}
\end{figure}

When the deformation parameters $z_-$ are finite, one
encounters a discrete choice of allowed contours that give rise to
finite but distinct results. To illustrate this point, consider the
simplest case of $R=1/2$ with $z_{-1}$ being the only non-vanishing
deformation parameter. We will also take $\mu$ to be real and
positive, and $t_{-1}$ to be real but have small negative imaginary
part. Then, $z$ is mostly imaginary with a small negative real
part. The asymptotic behavior of the integrand (\ref{integral1}) is
dominated by the $\exp(z_{-1} m^2)$ factor and converges when
integrated along contours that go to infinity along the unshaded
region, as illustrated in figure \ref{figb}.

The integral involving $H^+_q(am)$ can be performed along either of
the contours labeled by ${\bf A}$ and ${\bf B}$ in the figure. The
difference between the two contours will only contribute non-perturbatively in
the asymptotic expansion in small $t_{-1}/a^2 = t_{-1}/\mu^2 A^2$. For
a more general deformation, there could be many distinct contours
consistent with the perturbative expansion.  Therefore, in order to
define the $\tau$ function non-perturbatively, we need a definite
prescription for the contour.

One natural requirement is that the contour have a well defined limit
as the $t_{-n} \rightarrow 0$.  Let us assume, for simplicity, that
there is some $k>0$ such that $t_{-n}=0$ for $n>k$.\footnote{Such
deformations were also discussed in \cite{PandoZayas:2005tu}.}  Then,
the factor of $\exp[ z_-(m^{1/R})]$ is always oscillating along the
positive real axis. By adding a small imaginary part with appropriate
sign to $t_{-k}$, one can open a small region around the positive axis
for which the exponential factor $\exp[z_-(m^{1/R})]$ converges.
This is illustrated in figure \ref{figb}.  Both Henkel functions can
then be integrated over contours on the correct side of -- and
arbitrarily close to -- the real axis. In figure \ref{figb}, this
corresponds to choosing the contour ${\bf A}$.  We propose that this
contour is the correct non-perturbative definition of our integral,
(\ref{integral1}).  This prescription gives a natural, though not
unique, non-perturbative definition of the Toda flow. It would be
instructive if one could find some consistency check for this
prescription.

\subsection{Type 0B} 

Finally, let us briefly describe the extension to the case of type 0B
theory. This model turns out to be more subtle at the non-perturbative
level than the type 0A model.

We first consider the case of vanishing background RR flux. The model
is formulated as the quantum mechanics of an inverted harmonic
oscillator potential with fermions filling both sides of the potential
to the same level. There is an explicit $Z_2$ symmetry exchanging left
and the right side of the potential, and, as long as the unbounded
potential is regularized in a way that preserves the left/right
symmetry, the spectrum of states can be separated into sectors which
are odd and even with respect to this $Z_2$.  Since these two sectors
are decoupled from one another, the partition function factorizes into
a product of two $\tau$-functions \cite{Yin:2003iv},
\be Z(\mu) 
  = \tau_{odd}(z_{odd+}, z_{odd-},s) 
    \tau_{even}(z_{even+}, z_{even-},s). \ee
The Baker-Akhiezer functions for the odd and even sectors have an
integral transform,
\begin{align}
\Psi^s_{odd+}(\lambda_+) &= \int_0^\infty d\lambda_-^{-1} \,   
\sin(\lambda_+ \lambda^{-1}_-) \Psi_{odd-}^s (\lambda_-), 
\nonumber \\ 
\Psi^s_{even+}(\lambda_+) &= \int_0^\infty d\lambda_-^{-1} \,   
\cos(\lambda_+ \lambda^{-1}_-) \Psi_{even-}^s (\lambda_-).  
\label{sinecosinetransform}
\end{align}

Using the relation between $\tau$-functions and the Baker-Akhiezer
functions, one arrives at an integral expression of the form,
\begin{align}
\tau&_{even}(t_k,t_{-k},s)  = 
\nonumber\\
& 
 \int_0^\infty d M_i \,  
  {\Delta(M) \over \Delta( A)} 
  \prod_{i=1}^n (M_i A_i)^{\nu R + (R-1)/2} 
  \cosh (\nu A_i^R  M_i^R) \exp[ \nu t_-(M_i) -   n \log(M_i)] ,
\nonumber\\
\tau&_{odd}(t_k,t_{-k},s)  =  
\nonumber\\
&  \int_0^\infty d M_i  \,   
    {\Delta(M) \over \Delta (A)} 
    \prod_{i=1}^n (M_i A_i)^{\nu R + (R-1)/2} 
    \sinh (\nu A_i^R  M_i^R) 
    \exp[ \nu t_-(M_i) -   n \log(M_i)] \ ,  \label{0bim}
\end{align}
with 
\be t_k = - {A^{-k} \over \nu k} \ . \ee
These expressions are to be thought of as arising from the Fredholm
determinants, 
\be \tau_{even} = 
  \det(1 + e^{\beta \mu} \tilde{K}_{even-} \tilde K_{even+}), \qquad
\tau_{odd} = 
  \det(1 + e^{\beta \mu} \tilde{K}_{odd-} \tilde K_{odd+}),  \ee
where, for example,
\begin{align}
\label{deformedK3}
  [\tilde K_{even+} f](x_-) &=  
  \int [dx_+] \cos(e^{-i \beta} x_+ x_-) f(x_+), 
\nonumber \\
  [\tilde K_{odd+} f](x_+) &=  
  \int [dx_-] \sin(e^{-i \beta} x_+ x_-) f(x_-),
\end{align}
with  $[dx_\pm]$  defined in the same way as in (\ref{measurefactor}).

To relate these expressions to FZZT branes, observe that the 0B FZZT
determinant \cite{Takayanagi:2003sm,Takayanagi:2004jz,Seiberg:2004ei},
\be 
e^{\pm W(x)} = \det (X(x)^2- \mu_B)^{\pm 1} = \det \left((X_+(x) +
X_-(x))^2 - \mu_B\right)^{\pm 1},
\ee
asymptotes to 
\be \det\left(1 - {X_+^2 \over \mu_B'} \right)^{\pm 1} = e^{
\mp \sum_n {1 \over n} \mu_B'^{-n} X_+^{2n}} \
, \label{scaledloop3} \ee
which, for $R = 1/2$, matches the deformation of the $t_k$ implied by
the Baker-Akhiezer function. In a broad sense, we have therefore
succeeded in relating a scaling limit of FZZT amplitude to a
Kontsevich-like integral expression (\ref{0bim}). Note however that
unlike in the bosonic and the 0A cases, $R=1/2$ is not the special
radius for which (\ref{0bim}) can be re-expressed in matrix form using
the Harish-Chandra/Itzykson-Zuber formula.

Note, also, that the exponentiated macroscopic-loop operator perturbs
both the even and the odd sectors simultaneously. As such, it does not
probe the two sectors independently. A different object must be
introduced to generate purely even or odd deformations. It is not
clear what that object is.

To make sense of the integral expression, (\ref{0bim}), at the
non-perturbative level, it might seem that all one has to do is
address the issue of contours as we did for the 0A. However, there is
an additional subtlety in the 0B case.

To understand the origin of this subtlety, consider turning on a flux.
There are two proposals for the theory deformed by a flux.  One method
is to represent the flux by a relative shift between the chemical
potentials of the even and odd sectors. However, it was found that
such a prescription gives rise to violation of T-duality at a
non-perturbative level \cite{Gross:2003zz,Maldacena:2005he}.

An alternative prescription, where one adjusts the chemical potential
of the left and right-moving sectors independently, was proposed in
\cite{Maldacena:2005he}. This prescription leads to partition function
that is T-dual at the non-perturbative level. A second advantage of
this formalism is that there is a well defined notion of a fermi-sea,
which is lost when one deforms the even/odd sectors separately.

Unfortunately, the integrable structure of 0B has only been understood
in the case where we consider a flow which acts only on the even or
odd sectors separately.  In this formalism it is possible to turn on a
flux using the first method, in which the even and odd sectors have
different chemical potentials, but it seems very difficult to consider
the case where the left and right sectors are modified.  Moreover, by
splitting the fock space into even and odd sectors, one loses the
notion of a single Fermi-sea once any deformations are turned on.
This suggests that Toda flow given in \cite{Yin:2003iv} may not
represent a physical deformation of the background.

Perturbatively, none of this is an issue since the difference between
the left/right and even/odd bases only enters non-perturbatively.
However, to make the connection between FZZT-branes and Konsevich-like
integral expressions at the non-perturbative level, it seems one must
first gain a better understanding of the integrable structure of
0B.\footnote{Related discussion on integrable structure for minimal 0B
models can be found in \cite{Seiberg:2004ei}.}

\section{Conclusions}

In this paper, we found a precise relationship between the
Imbimbo-Mukhi model and the open string physics living on FZZT branes.
The key ingredient in this relationship, which did not play a role in
the $c<1$ theories, was our scaling limit.  Allowing the time
coordinate of the FZZT brane to be complex, we pushed the location of
the brane into the far Minkowski past.  This reduced the complicated
action of the full FZZT determinant to a pure Toda flow.

Using analytic continuation to reduce a brane amplitude to a purely
closed string process is reminiscent of the imaginary brane analysis
of \cite{Gaiotto:2003rm}.  There one does not require a limiting
procedure; rotating the brane to imaginary time is all that is
required to decouple the open string states.  This may well be true in
our story; however, the closed string states would, in general, be a
mixture of incoming and outgoing states.  The extra step of pushing
the brane to the far past is probably required to reduce the FZZT
brane to a purely incoming source.

We also considered the generalization to $\hat c=1$ models. For type
0A theory, we found that the integral expression generalizing the
Imbimbo-Mukhi formula \cite{Ita:2004yn,Hyun:2005fq} can be related to
the scaling limit of a macroscopic-loop operator provided the radius
is $R = (1/2) \sqrt{2 \alpha'}$. We also provided a prescription to
determine the $\tau$-function non-perturbatively along the integrable
flow.

A similar extension to 0B turned out to be more
subtle. Perturbatively, the computation is straightforward as it
reduces to two copies of the $c = 1$ theory.  These two copies
can be thought of as either the left and right wave functions or,
alternatively, the even and odd wave functions. 

Picking the even/odd basis, we find that, when $R = (1/2)\sqrt{2
\alpha'}$, the FZZT-brane determinant generates the appropriate Toda
flow.  Unfortunately, the flow is the same for the even and odd
sectors.  It is not clear if there is a better way to arrange the
FZZT-brane so that it only generates a flow in either the even or odd
sectors.  Perhaps a more serious problem is that even at $R =
(1/2)\sqrt{2 \alpha'}$, we do not find that the resulting integral can be
rewritten as a matrix integral.  This issue arises from the fact that
the integral is essentially two copies of the $c = 1$ integral, which
required $R = \sqrt{\alpha'}$ to be written in a matrix form.

Non-perturbatively, which basis one chooses matters.  In the case when
the flux vanishes, the even/odd basis seems most natural as the
problem continues to be manifestly factorizable.  One can define two
commuting Toda flows which generate separately even and odd
deformations.

However, when one turns on a flux, using the even/odd basis, one finds
that the 0B partition function is no longer T-dual to the partition
function of the 0A theory.  If, however, one uses the
left/right basis of \cite{Maldacena:2005he}, T-duality is restored.
Unfortunately, in the left/right basis, it is no longer obvious how one
should write down the Toda flow.  Thus, to properly generalize the
open/closed duality story to the 0B case appears to require a better
understanding of the integrable structure of 0B. We hope to address
this issue in the near future.

There are a number of interesting open issues. One is to explore
open/closed duality for Neumann FZZT branes. Unlike the
macroscopic-loop operator for the Dirichlet case \cite{Moore:1991sf},
the correct realization of the Neumann FZZT brane has not been
found. There is an interesting proposal for representing FZZT branes
using non-singlet sectors due to Gaiotto \cite{Gaiotto:2005gd}, but it
does not appear to reproduce the correct disk and annulus amplitudes
computed using world sheet techniques.

An understanding of the Neumann FZZT branes might shed some light on
our issues with the 0B analysis since the T-dual of the Dirichlet FZZT
branes in 0B are the Neumann branes in 0A.  Unlike in the Dirichlet
case, the Neumann brane acts as a source for winding modes. In the
matrix model language, winding modes are encoded by the holonomy of
the Wilson lines of the gauged quantum mechanics.  The fact that there
are two gauge groups in the 0A model is related to the existence of
the left and right sectors of the 0B theory.  Hence, if one can
understand the integrable structure of the winding modes of 0A in the
presence of flux, it might suggest the correct integrable structure
for the 0B side.

As a final point, we mention that the 0A and 0B models are simply two
isolated points among a class of non-critical string theories,
including affine type 0 theories and the type II theories.  These
theories are connected by a web of smooth deformations illustrated in
\cite{Seiberg:2005bx}.  It should be possible to explore the
integrable flow and T-duality of each of these theories at the
non-perturbative level, and to find a suitable generalization of the
Imbimbo-Mukhi integral.

\section*{Acknowledgments}

We would like to thank
T.~Eguchi,
D.~Gaiotto,
M.~Huang,
A.~Klemm,
J.~Maldacena,
D.~Shih, 
J.~Walcher,
X.~Yin, 
and especially  N.~Seiberg
for useful discussions.  AH would like to thank the Institute for
Advanced Study, the Taiwan National University, and the Sapporo Summer
Institute for hospitality where part of this work was done.  This work
was supported in part by the DOE grant DE-FG02-95ER40896 and funds
from the University of Wisconsin.

\appendix

\section*{Appendix A: Conventions for Toda Integrable Structure}
{\setcounter{section}{1}}
{\setcounter{equation}{0}}

In this appendix, we briefly review the machinery of the Toda
hierarchy in order to fix our conventions.  For a comprehensive
introduction to the Toda integrable hierarchy, see
\cite{TakasakiUeno,Takebe:1990fz,TakasakiBook}.

Following the analysis of \cite{Kostov:2002tk,KostovTalk},we can the
partition function of undeformed $c=1$ model in terms of the Fredholm
determinant,
\be Z(\mu) = \det(1+e^{- \beta \mu} K_- K_+), \ee
where
\begin{align} 
[K_+ f](x_-) &= \int dx_+ e^{i q^{-1} x_+ x_-} f(x_+)  ,
\nonumber \\
[K_- f](x_+) &= \int dx_- e^{-i  x_+ x_-} f(x_-) . 
\end{align}
Its $\log$ is therefore the sum over the eigenvalues of the operator,
\be \log(1 + e^{-\beta \mu} K_- K_+) \ . \ee
The orthonormal eigenstates, $|E \rangle$, can be expressed in the
position basis as
\be \psi_{0\pm}^E(x_\pm) = \langle x_\pm | E \rangle = {1 \over \sqrt{2 \pi}} e^{\mp i \phi_0(E)/2} x_\pm^{\pm i E - 1/ 2} \label{wavefunction1}, \ee 
and satisfy
\be \langle E | E' \rangle = \delta(E - E') \ . \ee
One then has
\be \log(Z(\mu)) =  \langle E| \log (1 + e^{-\beta (\mu+E)})| E \rangle 
=  \int dE \, \rho(E) \log (1 + e^{-\beta (\mu+E)}) \label{logZa}.
\ee
To compute $\rho(E)$, one imposes a boundary condition,
\be \psi^E_+(\Lambda) = \psi^E_-(\Lambda), \ee
which gives
\be \rho(E) = 
   - {1 \over 2 \pi} {\partial \over \partial E} \phi_0(E) 
   + {1 \over \pi}  \log \Lambda.
 \ee
The $\log(\Lambda)$ term non-universal and can be dropped. By
integrating by parts, one arrives at the standard expression,
\be \log(Z(\mu)) = -{\beta \over 2 \pi} \int dE  {\phi_0(E) \over 1 + e^{\beta(E+\mu)}} \ .  
\label{partintegral} 
\ee

In the deformed theory, the $\tilde K_\pm$  operator is modified;
\begin{align}
[\tilde K_+ f](x_-) &= \int [dx_+] e^{i
q^{-1} x_+ x_-} f(x_+),
\nonumber\\
[\tilde K_- f](x_-) &= \int [dx_-] e^{-i
 x_+ x_-} f(x_-) \ . 
\end{align}
The eigenvectors of the operator $\tilde K_- \tilde K_+$ can be expressed as
\be \tilde \psi^E_\pm(x_\pm) = e^{-i  U_\pm(x_\pm)} \psi^E_\pm(x_\pm), \ee
where 
\be \psi_\pm^E(x_\pm) 
  = \langle x_\pm | e^{\pm {i \over 2} \phi_0(E)} 
\hat {\cal W}_\pm^{-1} | E \rangle.
\ee
The operators $\hat {\cal W}_\pm$  are called ``dressing operators'' and take the form,
\be \hat {\cal W}_\pm = 
\hat W_\pm
e^{  \sum_{n \ge 1} z_{\pm n} \hat \omega^{\mp n/R}}, \qquad
\hat W_\pm 
 =
e^{\pm {1 \over 2} i  (1 \pm a) \phi(E)} 
e ^{-i  R \sum_{n \ge 1} v_{\pm n} \hat \omega^{\pm n/R} } ,
 \label{wform}
\ee
where
\be \hat \omega = e^{-i  \partial_E} \ . \ee
They satisfy the relation,
\be \hat {\cal W}_+ e^{-i \phi_0} \hat {\cal W}_-^{-1} = 1 \ . \label{biorthocond} \ee
It is the relation (\ref{biorthocond}), which, together with Theorem
1.5 of \cite{TakasakiUeno}, allows one to relate the computation of
the deformed matrix model to the Toda flow. The structure of the Toda
flow then allows one to compute $\phi(E)$ and the $v_{\pm n}$'s as a
function of $z_{\pm n}$'s, provided one specifies the initial
conditions satisfied by the undeformed theory. Once the $\phi(E)$ is
found, the partition function can be derived from
(\ref{partintegral}).

Since we already know the wave functions which diagonalize $K_- K_+$
in the undeformed theory, we can write the initial condition for $\hat
{\cal W}_\pm$ in the following form.
\be \hat {\cal W}_\pm(z_n = 0, z_{-n}=0) 
   = e^{\pm {1 \over 2} i(1 \pm a) \phi_0(E)} \ . \ee
Here, the parameter $a$ is introduced to allow for the ambiguity in
the overall phase of the wave functions which does not affect any
physical results. Indeed, different choices of $a$ give rise to the
same flow \cite{Takebe:1990fz}. We note this ambiguity here simply to
warn the reader of the different conventions found in the references,
with $a=0$ and $a=\pm 1$ being the most popular.

It is customary to parameterize $\hat W_\pm$ as 
\be \hat W_\pm = \sum_{j=0}^\infty \hat w_j^{\pm}(E,z_+,z_-) \hat
\omega^{\pm j}. \ee
Then, one can identify $\hat w_0^\pm$ as being related to $\phi(E)$ in
the parameterization of (\ref{wform}) according to
\be \hat w_0^\pm(E,z_+,z_-) = e^{\pm {i \over 2} (1 \pm a) \phi(E)} \ . \ee
This form of ${\cal W}$ can be matched to the convention of
\cite{TakasakiUeno} by identifying
\begin{align}
  \hat w_j^{({\infty \atop 0})}(s,z_+,z_-) &= w_j(E = i(s/R+1/2) , z_+, z_-) \ ,
\nonumber\\
\hat W^{({\infty \atop 0})}(z_+,z_-) &= \sum_{j=0}^\infty {\rm diag} 
 [ \hat w_j^{({\infty \atop 0})}] \Lambda^{\mp j}\ ,\cr
\Lambda &= \hat \omega^{-1/R} = e^{\partial_s}\ .  \label{dictionary}
\end{align}
With this choice, $\hat {\cal W}^{({\infty \atop 0})}$ takes the form
given in (1.2.9) and (1.2.10) of \cite{TakasakiUeno}. The sign and
phase of $z_{\pm n}$ was chosen in (\ref{deform}) in order for the
same $z_{\pm n}$ to appear on both side of this identification.

\section*{Appendix B: Baker Akhiezer Function}
{\setcounter{section}{2}}
{\setcounter{equation}{0}}

The Baker-Akhiezer function for the Toda lattice hierarchy is defined
in terms of the dressing operators
\cite{Takasaki:1993at};
\begin{align}
 \Psi^+_s(\lambda) & =
 \left(1 + \sum_{n=1}^\infty \hat w^{(\infty)}_n(s,z_+ ,z_- ) 
 \lambda^{-n/R}\right) \exp( z_+ (\lambda^{1/R}) + s \log \lambda^{1/R}) ,
\nonumber\\
\Psi^-_s(\lambda) & =\left(1 + \sum_{n=1}^\infty \hat w^{(0)}_n(s,z_+ ,z_- ) 
\lambda^{n/R}\right) \exp( z_- (\lambda^{-1/R}) + s \log \lambda^{1/R}) .
\end{align}
Equivalently, one has
\be 
\Psi^+_s(\lambda) = \hat {\cal W}^{(\infty)} \lambda^{s/R}, 
 \qquad  \Psi^-_s(\lambda) = \hat {\cal W}^{(0)} \lambda^{s/R} 
\ .\label{dressmono} 
\ee
Defining the Lax operator, $L_\pm$, and the Orlov-Shulman operators,
$M_\pm$, by
\be  L_\pm = \hat {\cal W}^{({\infty \atop 0})} \Lambda^{\pm R} 
 \hat  {\cal W}^{({\infty \atop 0}) -1},
\label{Lpm}
\ee
\be M_\pm = {1 \over R} {\cal W}^{({\infty \atop 0})}s 
 {\cal W}^{({\infty \atop 0}) \, -1},
\ee
it follows that the Baker-Akhiezer functions satisfy the differential
relations,
\be L_+ \Psi^+_s(\lambda) = \lambda \Psi^+_s(\lambda), \qquad 
 L_-  \Psi^-_s(\lambda) =   \lambda  \Psi^-_s(\lambda), \ee
\be M_+ \Psi^+_s(\lambda) =  
  \lambda {\partial \over \partial \lambda} \Psi^+_s(\lambda), \qquad 
 M_-  \Psi^-_s(\lambda)=
\lambda {\partial \over \partial \lambda}  \Psi^-_s(\lambda)\ . \ee 

These Baker-Akhiezer functions are intimately related to the fermion
wave functions of the double-scaled matrix model. Specifically,
\be (\psi^E_\pm(x_\pm))^*  =   
 \left. {1 \over \sqrt{2 \pi} } 
  x_\pm^{(1\mp 1)/2} \Psi^\pm_s(x_\pm^{\pm 1})   
\right|_{s =-  R( iE  + 1/2)}  .
\ee

Using the formal expression for the Baker-Akhiezer function,
(\ref{dressmono}), the orthogonality relation (\ref{biorthocond}), and the
dictionary (\ref{dictionary}), it follows that
\be{1 \over \sqrt{2 \pi} }\Psi^+_s(\lambda_+) =  
 \int d\lambda_-^{-1}\,  e^{i \lambda_+ \lambda_-^{-1}  }   
 {1 \over \sqrt{2 \pi} } \lambda_- \Psi^-_s(\lambda_-) .
\ee
The Fourier transform relation for the dual Baker-Akhiezer function
(\ref{fourier}), as well as the Bessel (\ref{besseltransform}) and the
sine/cosine transform (\ref{sinecosinetransform}) for the $\hat c=1$
models, can be derived along similar lines.

\section*{Appendix C: The Imbimbo-Mukhi integral for general $n$}
{\setcounter{section}{3}}
{\setcounter{equation}{0}}

\label{sec:MultipleFZZT}

In this section, we show how the properties of the Baker-Akhiezer
function determine the $(n+1)\times(n+1)$ Imbimbo-Mukhi integral in
terms of the $n\times n$ integral.  The computation is essentially the
same as the $1\times 1$ case, but the notation is a bit more
complicated. It is useful to define
\begin{equation}
  \epsilon_n = \sum_{i =1}^n \epsilon(\lambda_{+i}^{-1/R}).
\end{equation}
We then have the relation,
\begin{equation}
  \Psi_s^+ ( -\epsilon_{n},z_- ,\tilde{\lambda}_+)
 = \frac{
           \tau(-\epsilon_n - \epsilon(\lambda^{-1/R}),z_-,s)
        }
        {\tau(-\epsilon_n,z_-,s)}\,
 e^{-\sum_{i,n} \frac{1}{n} 
       \left( \frac{\tilde{\lambda}}{\lambda_i}\right)^{n/R} 
       + s\log(\tilde{\lambda}^{1/R})
   },
\end{equation}
so that
\begin{equation}
\label{shiftedtau}
  \tau(-\epsilon(\tilde{\lambda}^{-1/R}) - \epsilon_n,z_-,s)
=\Psi_s^+ ( -\epsilon_n,z_- ,\tilde{\lambda})
 \tau(-\epsilon_n,z_-,s)\,
e^{\sum_{i,n} \frac{1}{n} 
       \left( \frac{\tilde{\lambda}}{\lambda_i}\right)^{n/R} 
       - s\log(\tilde{\lambda}^{1/R})
  }.
\end{equation}
Next, note that
\begin{align}
   \Psi_s^+& ( -\epsilon_n,z_- ,\tilde{\lambda}_+) 
\\
&= \int
 d\tilde{\lambda}_-^{-1} e^{i \tilde{\lambda}_+ \tilde{\lambda}_-^{-1}}
\frac{1}{\sqrt{2\pi}} 
\Psi_s^- (-\epsilon_n,z_- ,\tilde{\lambda}_-) \tilde{\lambda}_-
  \cr
&=\int
 d\tilde{\lambda}_-^{-1} e^{i \tilde{\lambda}_+ \tilde{\lambda}_-^{-1}}
\frac{1}{\sqrt{2\pi}}
\frac{\tau(-\epsilon_n,z_- 
        - \epsilon(\tilde{\lambda}_-^{1/R}),s+1)} 
     {\tau( - \epsilon_n,z_-,s)}
\exp [z_- (\tilde \lambda_-^{-1/R}) + (s+R) \log(\tilde \lambda_-^{1/R})]. \nonumber 
\end{align}
Substituting this into (\ref{shiftedtau}) gives
\begin{multline}
   \tau(- \epsilon_n-\epsilon(\tilde{\lambda}^{-1/R}_+),z_-,s)
=
    e^{\sum_{i,n} \frac{1}{n} 
        \left(\tilde{\lambda}_+/\lambda_i\right)^{n/R} 
       - s\log(\tilde{\lambda}^{1/R}_+)
    }
\cr
\times
\int
 d\tilde{\lambda}_-^{-1} e^{i \tilde{\lambda}_+ \tilde{\lambda}_-^{-1}}
 \frac{1}{\sqrt{2\pi}}
  \tau(-\epsilon_n,z_- 
        - \epsilon(\tilde{\lambda}_-^{1/R}),s+1)
   \exp [z_- (\tilde \lambda_-^{-1/R}) 
    + (s+R) \log(\tilde \lambda_-^{1/R})],
\end{multline}
or, equivalently, using the notation, (\ref{IMnotation}), and shifting
$n \to n-1$, we find the recursion relation,
%
% \beq
% \label{IMrecursion}
% \lefteqn{   \tau(- \sum_{i=1}^{n} \epsilon(\mu^{1/2R} A_i),z_-,s)
% = \mu^{-(s+1)/R}
%   \left( \prod_i \frac{1}{A_i - A_{n}}\right) e^{-s \log(A_{n})+ \sum_i \log(A_i)}}
% \\
% &&\times 
%   \int \frac{d M_{n}}{\sqrt{2\pi}}\, e^{-\nu M^R_{n} A^R_{n}}
% \tau(-\sum_{i=1}^{n-1} \epsilon(\mu^{1/2R} A_i),z_- - \epsilon( \mu^{-1/2R}M^{-1}_{n}),s+1)
% e^{-\nu t_-(M_{n}) - (s+1) \log(M_{n})}. \nonumber 
% \eeq
\beq
\label{IMrecursion}
\lefteqn{   Z_n(A,t_-,s)
= 
  \left( \prod_{i=1}^{n-1} \frac{1}{A_i - A_{n}}\right) e^{-s \log(A_{n})+ \sum_{i=1}^{n-1} \log(A_i)}}
\\
&&\times 
  \int \frac{d M_{n}}{\sqrt{2\pi}}\, e^{-\nu M^R_{n} A^R_{n}}
Z_{n-1}(A,t_- - \epsilon( M^{-1}_{n}),s+1)
e^{-\nu t_-(M_{n}) - (s+1) \log(M_{n})}. \nonumber 
\eeq
It is straightforward to check, by
direct substitution, that (\ref{notselfdual}) solves (\ref{IMrecursion}).

\bibliography{ceq1}\bibliographystyle{utphys}

\end{document}